\documentclass[screen,balance=false,sigconf]{acmart}

\copyrightyear{2019} 
\acmYear{2019} 
\setcopyright{acmcopyright}
\acmConference[KDD'19]{The 25th ACM SIGKDD Conference on Knowledge Discovery and Data Mining}{August 4--8, 2019}{Anchorage, AK, USA}
\acmBooktitle{The 25th ACM SIGKDD Conference on Knowledge Discovery and Data Mining (KDD'19), August 4--8, 2019, Anchorage, AK, USA}
\acmPrice{15.00}
\acmDOI{10.1145/3292500.3330726}
\acmISBN{978-1-4503-6201-6/19/08}
\usepackage{soul}
\usepackage{url}
\usepackage[utf8]{inputenc}
\usepackage{caption}
\usepackage{cleveref}
\usepackage{xcolor}

\usepackage{amsmath}
\usepackage{amssymb}
\usepackage{wasysym}
\usepackage{graphicx}
\usepackage{balance}
\usepackage{appendix}
\graphicspath{{figure/}}
\usepackage[colorinlistoftodos]{todonotes}

\usepackage{algorithm}
\usepackage{algorithmicx}
\usepackage{algpseudocode}

\usepackage{booktabs}
\usepackage[inline]{enumitem}
\usepackage{enumitem}

\usepackage{tikz}
\usetikzlibrary{arrows}
\usetikzlibrary{shapes.geometric}

\newcommand{\ignore}[1]{}

\newcommand{\model}{s$^2$Meta}
\newcommand{\smodel}{\emph{s$^2$Meta}\space}

\newcommand{\myvec}[1]{\boldsymbol{{#1}}}
\newcommand{\mymat}[1]{\boldsymbol{{#1}}}

\newcommand{\tabincell}[2]{\begin{tabular}{@{}#1@{}}#2\end{tabular}}

\newcommand{\userset}{\mathcal{U}}
\newcommand{\itemset}{\mathcal{I}}
\newcommand{\scenarioset}{\mathcal{C}}

\newcommand{\metatrainset}{\mathbb{T}_{\text{meta-train}}}
\newcommand{\metatestset}{\mathbb{T}_{\text{meta-test}}}
\newcommand{\trainset}{D^{\text{train}}}
\newcommand{\testset}{D^{\text{test}}}
\newcommand{\trainloss}{\mathcal{L}}
\newcommand{\testloss}{\mathcal{L}^{\text{test}}}

\newcommand{\metaparam}{\myvec{\omega}}
\newcommand{\nmetaparam}{\omega}
\newcommand{\nupdateparam}{\omega_u}
\newcommand{\updateparam}{\myvec{\omega}_u}
\newcommand{\ninitparam}{\omega_R}
\newcommand{\initparam}{\myvec{\omega}_R}
\newcommand{\nstopparam}{\omega_s}
\newcommand{\stopparam}{\myvec{\omega}_s}
\newcommand{\nrecomparam}{\theta}
\newcommand{\recomparam}{\myvec{\nrecomparam}}

\newcommand{\forgetgate}{\beta}
\newcommand{\inputgate}{\alpha}

\def \xiaowei #1{{\color{orange}{\textbf{[Xiaowei: #1]}}}}

\begin{document}

\title{Sequential Scenario-Specific Meta Learner for \\ Online Recommendation}
%Few-shot Scenario-Aware Recommender Systems based on AutoML}
%\author{Zhengxiao Du}
%\affiliation{%
%  \institution{Tsinghua University}
%}
%\email{duzx16@mails.tsinghua.edu.cn}
%
%\author{Xiaowei Wang}
%\affiliation{%
%  \institution{Alibaba Group}
%}
%\email{daemon.wxw@alibaba-inc.com}
%
%\author{Jie Tang}
%\affiliation{%
%  \institution{Tsinghua University}
%}
%\email{jietang@tsinghua.edu.cn}
%
%\author{Jingren Zhou}
%\affiliation{%
%  \institution{Alibaba Group}
%}
%\email{jingren.zhou@alibaba-inc.com}
%
%\author{Hongxia Yang}
%\affiliation{%
%  \institution{Alibaba Group}
%}
%\email{yang.yhx@alibaba-inc.com}

\author[Z. Du*, X. Wang*, Y. Zhang, H. Yang, J. Zhou, J. Tang]{
    Zhengxiao Du$^{1}$, Xiaowei Wang$^{2}$,  Hongxia Yang$^{2}$, Jingren Zhou$^{2}$, Jie Tang$^{1}$
}
\affiliation{
    $^1$ Department of Computer Science and Technology, Tsinghua University
}
\affiliation{
    $^2$ DAMO Academy, Alibaba Group
}
\email{
  duzx16@mails.tsinghua.edu.cn,
  {daemon.wxw, yang.yhx, jingren.zhou}@alibaba-inc.com,
   jietang@tsinghua.edu.cn
}

\renewcommand{\shortauthors}{Du et al.}

\sloppy

% The abstract is a short summary of the work to be presented in the article.
\begin{abstract}
Cold-start problems are long-standing challenges for practical recommendations. Most existing recommendation algorithms rely on extensive observed data and are brittle to recommendation scenarios with few interactions. This paper addresses such problems using \textit{few-shot learning} and \textit{meta learning}. Our approach is based on the insight that having a good generalization from a few examples relies on both a generic model initialization and an effective strategy for adapting this model to newly arising tasks. 
To accomplish this, we combine the scenario-specific learning with a model-agnostic sequential meta-learning and unify them into an integrated end-to-end framework, namely \textbf{S}cenario-specific \textbf{S}equential \textbf{Meta} learner (or \smodel). By doing so, our \textit{meta-learner} produces a generic initial model through aggregating contextual information from a variety of prediction tasks while effectively adapting to specific tasks by leveraging learning-to-learn knowledge. Extensive experiments on various real-world datasets demonstrate that our proposed model can achieve significant gains over the state-of-the-arts for cold-start problems in online recommendation\footnote{The source code is available at \url{https://github.com/THUDM/ScenarioMeta}}. Deployment is at the Guess You Like session, the front page of the Mobile Taobao; and the illustration video can also be watched from the link\footnote{\url{https://youtu.be/TNHLZqWnQwc}}. 
\end{abstract}



%
% Keywords. The author(s) should pick words that accurately describe the work being
% presented. Separate the keywords with commas.

\maketitle

\section{Introduction}
The personalized recommendation is an important method for information retrieval and content discovery in today’s information-rich environment. Personalized recommender systems, where the recommendation is generated according to users' past behaviors or profiles, have been proven effective in domains including E-Commerce \cite{Amazon:LindenSY03}, social networking services \cite{Twitter:KyweLZ12}, video-sharing websites \cite{Youtube:CovingtonAS16}, among many others. Traditionally, a personalized recommender system can be seen as a mapping $\mathcal{U}\times\mathcal{I}\rightarrow\mathbb{R}$, where $\mathcal{U}$ is the user set and $\mathcal{I}$ is the item set. The mapping result can be a real value for explicit ratings or a binary value for implicit feedback \cite{NeuralMF:HeLZNHC17,MFforRS:KorenBV09,Caser:TangW18,MFforRS:KorenBV09}. This setting usually assumes that the behavior pattern of the same user is relatively stationary in different contexts, which is not true in many practical tasks \cite{DecisionMaking:PAYNE1992107,BrandChoice}. For example, during the Singles Day Promotion (Double 11) period, the largest online shopping festival in China, consumers sometimes shop impulsively allured by the low discounts. In such scenarios, the contextual information of Double 11 is quite critical. It also has been shown that including contextual information leads to better predictive models and better quality of recommendations \cite{MultiContext:AdomaviciusSST05,CustomerContext:PalmisanoTG08}.

Though context-aware recommender systems have been proven effective \cite{Context-aware:AdomaviciusT15}, they are facing several challenges. Firstly, a large portion of scenarios in a system is actually long-tailed, without enough user feedback.  Moreover, the life cycle of a scenario can be quite short. In Taobao, most promotions end within a few days or even a few hours after being launched, without enough time to collect sufficient user feedback for training. Therefore, training scenario-specific recommenders with limited observations are practical requirements. Most existing recommendation algorithms rely on extensive observed data and are brittle to new products and/or consumers \cite{MetaColdStart:VartakTMBL17,DropoutNet:VolkovsYP17}. Although the cold-start problem can be tackled by cross-domain recommender systems with domain knowledge being transferred, they still require a large amount of shared samples across domains \cite{EMCDR:Man,CoNet:HuZY18,CMF:SinghG08}. Secondly, when training a predictive model on a new scenario, the hyperparameters often have a great influence on the performance and optimal hyperparameters in different scenarios may differ significantly \cite{HyperImportance:AshrafiSDABPM15,HyperImportance:RijnH18}. Finding a right combination of hyperparameters usually requires great human efforts along with sufficient observations. 

This paper addresses the problem of cold-start scenario recommendation with the recent progress on \textit{few-shot learning} \cite{MatchingNetwork:Vinyals,ProtoNet:Snell,oneshotrelational:Xiong} and \textit{meta-learning} \cite{MAML:Finn,Meta-LSTM:Ravi,Reptile:abs-1803-02999,GradientByGradient:AndrychowiczDCH16}. Our approach is to build a \emph{meta learner} that learns how to instantiate a recommender with good generalization capacity from limited training data. The framework, which generates a scenario-specific recommender by a sequential learning process, is called \textbf{S}cenario-specific \textbf{S}equential \textbf{Meta} learner (or \smodel). The sequential process, which resembles the traditional machine learning process controlled by human experts, consists of three steps: the meta learner automatically initializes the recommender to be broadly suitable to many scenarios, finetunes its parameters with flexible updating strategy, and stops learning timely to avoid overfitting. The policy of meta learner is learned on a variety of existing scenarios in the system and can guide the quick acquisition of knowledge in new scenarios. By automating the learning process in each scenario, we can also reduce the human efforts needed to train the predictive model on new scenarios that continuously appear in the recommender systems.

\noindent\textbf{Organization.} The remainder of this paper is organized as follows: in \Cref{sec:related_work}, we review related works. \Cref{sec:preliminary} introduces the problem definition and meta learning settings. \Cref{sec:approach} gives a detailed description of the proposed \smodel framework. Experimental results on large-scale public datasets and the newly released Taobao theme recommendation dataset   are shown in \Cref{sec:experiment}. At last, \Cref{sec:conclusion} concludes the paper.

\section{Related Work} \label{sec:related_work}
In this section, we go over the related works on context-aware recommendation, cross-domain recommendation, and meta learning respectively. 
\subsection{Context-Aware Recommendation}
The importance of contextual information has been recognized by researchers and practitioners in many disciplines, including E-Commerce personalization \cite{CustomerContext:PalmisanoTG08}, information retrieval \cite{IR:Ingwersen2005}, data mining \cite{DataMining:BerryL97} and marketing \cite{DecisionMaking:PAYNE1992107}, among many others \cite{Database:StefanidisPV07,BrandChoice}. Relevant contextual information does matter in recommender systems, and it is important to take account of contextual information, such as time, location, or acquaintances' impacts \cite{NextGenerationRS:AdomaviciusT05,MultiContext:AdomaviciusSST05,CustomerContext:PalmisanoTG08}. Compared to traditional recommender systems that make predictions based on the information of users and items, context-aware recommender systems \cite{MultiContext:AdomaviciusSST05,Context-aware:AdomaviciusT15} make predictions in the space of $\mathcal{U}\times\mathcal{I}\times\mathcal{C}$, where $\mathcal{C}$ is the context space.
The contextual information can be observable (e.g., time, location, etc), or unobservable (e.g., users' intention). The latter case is related to \emph{Session-based Recommendation} \cite{SessionBased:HidasiK18} or \emph{Sequence-aware Recommendation} \cite{SequenceAware:QuadranaCJ18}. Even if the contextual information is fully-observable, the weight of different types of information is entirely domain-dependent and quite tricky to be tuned for cold-start scenarios.  

\subsection{Cross-Domain Recommendation}
Traditional recommender systems suggest items belonging to a single domain and this is not perceived as a limitation, but as a focus on a particular market \cite{MFforRS:KorenBV09,Youtube:CovingtonAS16,Twitter:KyweLZ12}. However, nowadays, users provide feedback for items of different types, express their opinions on different social media and different providers. Providers also wish to cross-sell various categories of products and services,  especially to  new users. Unlike traditional recommender systems that work on homogeneous users and items, cross-domain recommender systems \cite{WhenRSFail:EkstrandR12,CrossDomain:CantadorFBC15}, closely related to Transfer Learning \cite{TransferL:Torrey}, try to combine the information from heterogeneous users and/or items. In \cite{CDCF:LoniSLH14,CMF:SinghG08}, they extend the traditional matrix factorization \cite{MFforRS:KorenBV09} or factorization machines \cite{FM:Rendle10} with interacted information from an auxiliary domain to inform recommendations in a target domain. In \cite{EMCDR:Man}, a multi-layer perceptron is used to capture the nonlinear mapping function across different domains. In \cite{CoNet:HuZY18}, they propose a novel neural network CoNet, with architecture designed for knowledge transfer across domains. However, all these methods require a large amount of interacted data of the shared users or items between the source	and target domains.  In \cite{HetergeneousDomain:YangHQ}, they propose to align the different features in two domains without shared users or items with semantic relatedness. However, they assume that the relatedness of features in the source and target domains can be inferred from domain prior knowledge. 

\subsection{Meta Learning}\label{subsec:meta learning}
Meta learning \cite{MetaLearning:Brazdil2008}, or \emph{Learning to learn}, which aims to learn a learner that can efficiently complete different learning tasks, has gained great success in few-shot learning settings \cite{MatchingNetwork:Vinyals,ProtoNet:Snell,Meta-LSTM:Ravi,MAML:Finn}. The information learned across all tasks guides the quick acquisition of knowledge within each separate learning task. Previous work on meta learning can be divided into two groups. One is the metric method that learns a similarity metric between new instances and instances in the training set. Examples include Siamese Network \cite{SiameseNetwork:koch2015}, Matching Network \cite{MatchingNetwork:Vinyals}, and Prototypical Network \cite{ProtoNet:Snell}. The other is the parameter method that directly predicts or updates the parameters of the classifier according to the training data. Examples include MAML \cite{MAML:Finn}, Meta-LSTM \cite{Meta-LSTM:Ravi} and Meta Network \cite{MetaNetwork:MunkhdalaiY17}. Most methods follow the framework in \cite{MatchingNetwork:Vinyals}, trying to solve a group of learning tasks of the same pattern, e.g., learning to recognize the objects of different categories with only one example per category given \cite{OneShot:Fei-FeiFP06}. Previous works on the application of \emph{learning to learn} in recommender systems mainly focus on the algorithm selection \cite{WhenRSFail:EkstrandR12,SelectMeta:CunhaSC16,MetaCF:CunhaMRS}, leaving other parts of the learning process less explored. In \cite{MetaColdStart:VartakTMBL17}, they propose to use meta learning to solve the user cold-start problem in the Tweet recommendation. However, they use a neural network to directly output parameters of the recommender, which limits the representation ability of their model.  
\section{Preliminary}
\label{sec:preliminary} 

\begin{figure*}[!ht]
\centering
%\begin{tikzpicture}[scale=1]
%   \node [anchor=south] at (5,8.5) {$\theta_{t+1}=\myvec{\forgetgate}_t\odot\theta_t - \myvec{\alpha}_t\odot\nabla_{\theta_t}\mathcal{L}_t$};
%   \node (ml) [rectangle, very thick, draw=black, minimum size=10mm] at (5, 8) {Meta Learner $M^\omega$};
%	\node (r) [rectangle, very thick, draw=black, minimum size=10mm] at (5, 5) {Recommender $R^\theta$};
%	\path (4.5, 5.6) edge[->] node[left] {$\theta_t$, $\mathcal{L}_t$, $\triangledown_{\theta_t}{\mathcal{L}_t}$} (4.5, 7.4);
%	\path (6, 7.4) edge[->] node[left] {$\theta_{t+1}$} (6, 5.6);
%	\path (8.5, 5) edge[->, thick] node[above] {$B_t$} (6.5,5);
%\end{tikzpicture}
\includegraphics[width=0.88\textwidth]{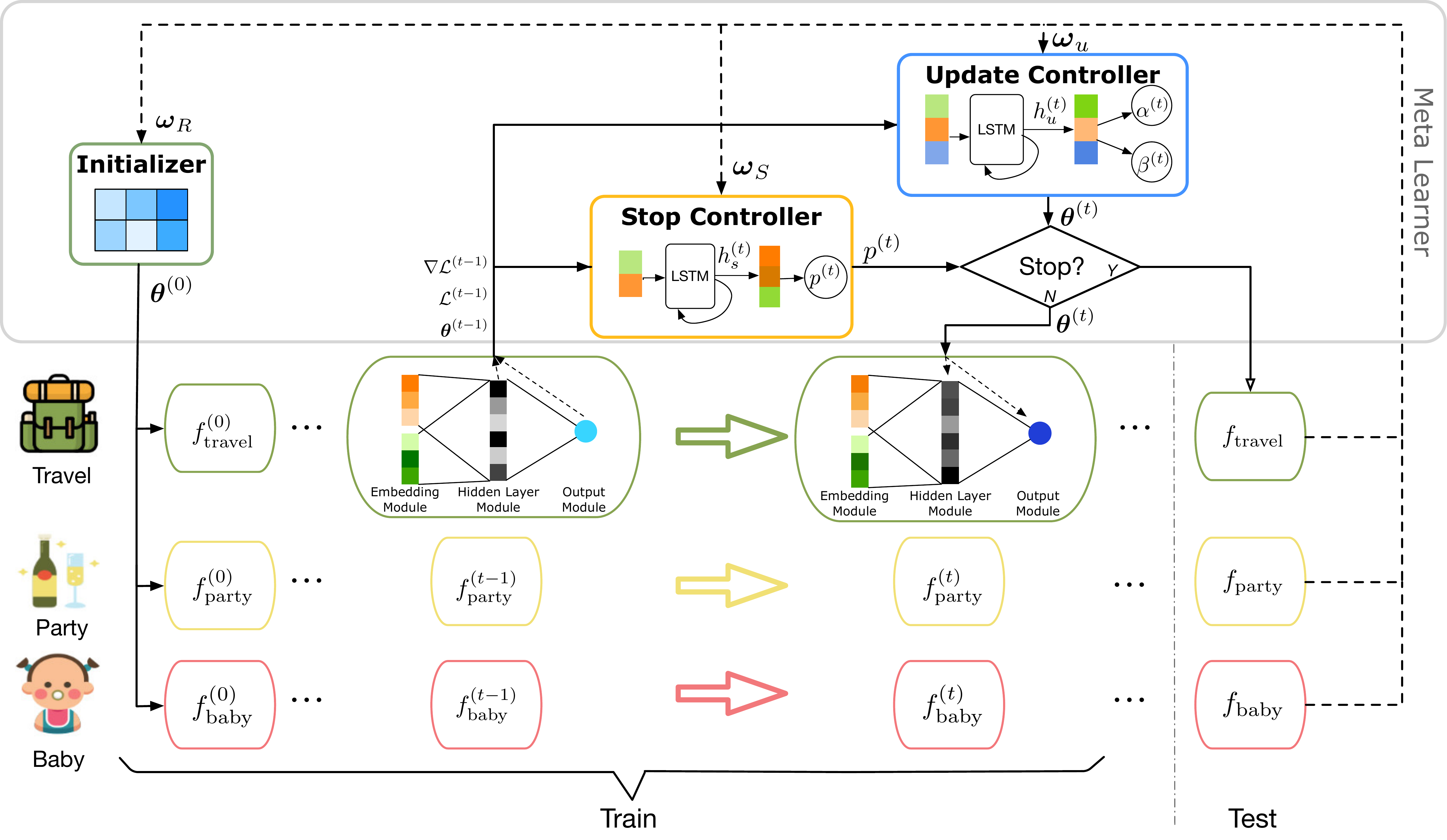}
\vspace{-0.13in}
\caption{The framework of meta learner and recommender. In each scenario, the recommender is initialized by the initializer and updated by the update controller. After the learning process is stopped by the stop controller, the loss of the final recommender on the test set is computed, and the parameters of meta learner are updated by the meta-gradient.}
\end{figure*}
\vspace{-0.05in}

In this section, we formally define the problem of scenario-aware recommendation and introduce the meta learning settings. We summarize the notations in  \Cref{tab:notation}.

\subsection{Problem Formulation}
A scenario-based recommender system is represented as $\{\userset,\itemset,\scenarioset\}$, where $\userset$ is the user set, $\itemset$ is the item set and $\scenarioset$ is the scenario set. A scenario $c \in \scenarioset$ is defined as a common context in which users interact with items and $(u,i)\in \mymat{H}_c$ represents that user $u$ interacted with item $i$ in the scenario $c$. Each scenario $c$ is connected with a recommender function $f_c: \userset\times\itemset\rightarrow\mathbb{R}$, where $f_c(u,i)$ is the ranking score of item $i$ for user $u$ in scenario $c$. For each scenario $c$, we have access to a training set $\trainset_{c}=\{(u_k,i_k)_{k=1}^{N_c}\}$ where $(u_k,i_k)\in \mymat{H}_c$. Our task is to recommend top-n items for each user $u$ in scenario $c$ and maximize the probability of the user's follow-up behaviors, e.g., clicks or conversions. Notice that, compared to previous works in \textit{cross-domain recommender systems} that require a large amount of training samples on each domain, our work studies the case where the size of $\trainset_c$ is limited. The latter usually happens in practice, e.g., new promotion scenarios come out with very sparse related user-item instances.  

\begin{table}[!ht]
	\caption{Notations}
	\centering
	\label{tab:notation}
	\begin{tabular}{c|p{2.2in}}
		\hline
		Notation & Definition or Descriptions\\
		\hline
		\hline
		$\userset, \itemset, \scenarioset$ & the user set, item set and scenario set\\
		\hline
		$m=|\userset|,n=|\itemset|$ & numbers of users and items \\
		\hline
		$c$ & a recommendation scenario\\
		\hline
		$\mymat{H}_c\subset\userset\times\itemset$ & the interaction set of $c$\\
		\hline
		$T_c$ & the learning task connected with $c$  \\
		\hline
		$\trainset_c, \testset_c$ & the training set and testing set of $T_c$\\
		\hline
		$f_c: \userset\times\itemset\rightarrow\mathbb{R}$ & the recommender function for $c$\\
		\hline
		$\recomparam_c$ & parameters of $f_c$\\
		\hline
		$\mymat{U},\mymat{I}$ & embedding matrices for users and items\\
		\hline
		$M$ & meta learner\\
		\hline
		$\metatrainset,\metatestset$ & meta-training set and meta-testing set \\
		\hline
		$\metaparam=\{\initparam,\updateparam,\stopparam \}$ & parameters of the meta learner \\
		\hline
		$\initparam$ & shared initial parameters for $f_c$\\
		\hline
		$\updateparam,\stopparam$ & parameters of update and stop controllers\\
		\hline
		$\myvec{\inputgate}, \myvec{\forgetgate}$ & input gate and forget gate \\
		\hline
		$p^{(t)}$ & stop probability at step $t$\\
		\hline
		\end{tabular}
\end{table}

\subsection{Scenario-Specific Meta Learning Settings}
The recommendation task in the scenario $c$ can be considered as a learning task, whose training set is $\trainset_c$. Our goal is to learn a meta learner $M$ that, given $\trainset_c$, predicts parameters of $f_c$, $\recomparam_c$.

Similar to the standard few-shot learning settings \cite{MatchingNetwork:Vinyals,Meta-LSTM:Ravi}, we assume access to a set of training tasks as the meta-training set, denoted as $\metatrainset$. Each training task $T_c\in \metatrainset$ corresponds to a scenario $c$ and has its training/testing pairs: $T_c=\{\trainset_c, \testset_c\}$. $\testset_c$ is the set of pairwise testing instances: $\testset_c:=\{(u,i,i^-)|(u,i)\in \mymat{H}_c\wedge (u,i)\notin\trainset_c\wedge (u,i^-)\notin\mymat{H}_c\}$. The meta-training set can be easily built with the previous scenarios in the recommender system, by randomly dividing the user-item interactions in each scenario into the training set and testing set. The parameters $\metaparam$ of the meta learner is optimized w.r.t. the meta-training objective:

\begin{equation}
\label{eqn:objective}
\min_{\myvec{\omega}}{\mathbb{E}_{T_c}}\big[\mathcal{L}_\omega(\testset_c|\trainset_c)\big],
\end{equation}

\noindent where $\mathcal{L}_{\nmetaparam}(\testset_c|\trainset_c)$ is the loss on the testing set $\testset_c$ given the training set $\trainset_c$ and meta learner parameters $\metaparam$. Specifically, we use the hinge loss as the loss function:
\begin{align}
\label{eqn:test_loss}
	 \mathcal{L}_{\metaparam}(\testset_c|\trainset_c) &= \sum_{(u_k,i_k,i_k^-)\in \testset_c}{\frac{\ell(u_k,i_k,i_k^-;\recomparam_c)}{|\testset_c|}},\\
	 \label{eqn:loss_function}
	\ell(u,i, i^-;\recomparam_c)&=\max\big(0, \gamma-f(u,i;\recomparam_c)+f(u,i^-;\recomparam_c)\big),
\end{align}

\noindent where the margin $\gamma$ is set to 1 and $\recomparam_c=M(\trainset_c;\metaparam)$. $\recomparam_c$ is generated via a sequential process, which we call \emph{scenario-specific learning}. During the scenario-specific learning, the meta learner initializes $\recomparam_c$ and updates it via gradient descent of flexible steps. After the learning, we will dump the training set and use the recommender $f_c$ for further tasks in the scenario $c$. During the evaluation, a different set of learning tasks is used, called meta-testing set $\metatestset$, whose scenarios are unseen during meta-training, i.e. $\metatrainset\cap\metatestset=\emptyset$.

\section{The Proposed Framework}
\label{sec:approach}
In this section, we detail the modules of the recommender network and meta learner for scenario-specific learning.

%\iffalse
%\subsection{Scenario-specific Learning}
%Given a training set $\trainset_c$ in the scenario $c$, its recommender  is instantiated from $\trainset_c$ via the learning process, which we call \emph{scenario-specific learning}. The meta learner initializes the recommender and update the scenario-specific parameters $\recomparam^c$ via the multistep gradient descent. At each step, a batch $B=\{u_n, i_n, i_n^-\}_{n=1}^N$ with the fixed size $N$ is sampled from $\trainset_c$, where $(u_n, i_n)\in\trainset_c$ and $i_n^-$ is sampled from $\mathcal{I}$ such that $(u_n, i_n^-)\notin\trainset_c$. We define  the following hinge loss function as the scenario-specific learning:
%\begin{equation}
%	\ell(\theta_c, u,i, i^-)=\max\big(0, \gamma-\hat{f}_c (\theta_c, u,i)+\hat{f}_c(\theta_c, u,i^-)\big), 
%\end{equation}
%where the margin $\gamma$ is set to 1. After the scenario-specific learning, we will dump the training set and use the recommender $\hat{f}_c$ for further tasks. 
%\fi

\subsection{Recommender Network}
\label{subsec:recommender}
We apply a feedforward neural network as the recommender $f_c$, which consists of three modules, to take inputs $(u,i)$ and generate corresponding outputs $f_c(u, i)$.

\textbf{Embedding Module ($u,i\rightarrow \myvec{e}_u, \myvec{e}_i$)}: This module consists of two embedding matrices $\mymat{U}\in\mathbb{R}^{m\times d}$ and $\mymat{I}\in\mathbb{R}^{n\times d}$ for users and items respectively. The embeddings can be generated from user/item attributes and/or general interactions without contextual information, with methods in collaborative filtering \cite{MFforRS:KorenBV09,BPR:RendleFGS09} or network embedding \cite{DeepWalk:PerozziAS14,GraphSage:HamiltonYL17}. The user $u$ and item $i$ are first mapped to one-hot vectors $\myvec{x}_u\in\{0,1\}^{m}$ and $\myvec{x}_i\in\{0,1\}^{n}$, where only the element corresponding to the user/item id is 1 and all others are 0. The one-hot vectors are then transformed into continuous representations by embedding matrices: $\myvec{e}_u=\mymat{U}\myvec{x}_u $ and $\myvec{e}_i=\mymat{I}\myvec{x}_i$.

\textbf{Hidden Layer Module ($\myvec{e}_u,\myvec{e}_i\rightarrow \myvec{z}_{ui}$)}: This module is the central part of the recommender. Similar to deep recommendation models in   \cite{WideRS:Cheng0HSCAACCIA16,Youtube:CovingtonAS16}, the user and item embeddings are concatenated as $\myvec{e}_{ui}=[\myvec{e}_u,\myvec{e}_i]$. $\myvec{e}_{ui}$ is then mapped by $L$ hidden layers to a continuous representation of user-item interaction $\myvec{z}_{ui}=\phi_L(\cdots(\phi_1(\myvec{e}_{ui}))\cdots)$. The $l$-th layer can be represented as:

\begin{equation}
	\phi_l(\myvec{z})=\operatorname{ReLU}(\mymat{W}_{l}\myvec{z}+\myvec{b}_{l}),
\end{equation}

\noindent where $\mymat{W}_{l}$ and $\myvec{b}_{l}$ are the weight matrix and the bias vector of the $l$-th layer.

%\textbf{Scenario Mapping Module ($\myvec{e}_u,\myvec{e}_i\rightarrow \myvec{z}_u,\myvec{z}_i$)}: This module is the main part of recommender. Unlike previous deep recommendation models  \cite{WideRS:Cheng0HSCAACCIA16,Youtube:CovingtonAS16}, in which the user and item embeddings are concatenated at the early age, we use two Multilayer Perceptrons (MLPs) to map the user and item embeddings respectively. This method is similar to the entity mapping in TransH\cite{TransH:WangZFC14} and TransR\cite{TransR:LinLSLZ15}, where users and items are identical to head and tail entities, scenarios to relations. Take the mapping of user embeddings as example, the continuous representation is mapped by $L$ layers as $\myvec{z}_u=\phi_L(\cdots(\phi_1(\myvec{e}_u))\cdots)$. The mapping of item representations is computed in the same way.

\textbf{Output Module  $\myvec{z}_{ui}\rightarrow f_c(u, i)$}: This module computes the recommendation score $f_c(u, i)$ based on the mapped representation of user-item interaction from the last hidden layer. This is achieved by a linear layer as $f_c(u, i)=\myvec{w}^T\myvec{z}_{ui}$, where $\myvec{w}$ is the weight of output layer.

The recommender parameters $\recomparam_c$, include $\{(\mymat{W}_l,\myvec{b}_l)_{l=1}^L, \myvec{w}\}$, are learned during scenario-specific learning. Note that the embedding matrices $\mymat{U}$ and $\mymat{I}$ are not included, to improve the recommender's generalization ability for unobserved users and items in $\trainset_c$. In other words, embedding matrices are shared across different scenarios and kept fixed in scenario-specific learning. 

\subsection{\smodel}
\label{subsec:meta_learner}
\begin{algorithm}[tb]
\caption{Training Algorithm of Meta Learner}
\label{algo:meta_train}
\begin{algorithmic}[1]
	\Require Meta-training set $\metatrainset$, \ignore{Update Controllers $M_u$ with parameters $\updateparam$, Stop Controller $M_s$ with parameters $\stopparam$, \xiaowei{$\initparam$ ?}}\ignore{Recommender $f$ with parameters $\recomparam$, }Loss function $\mathcal{L}$
	\State $\updateparam,\stopparam,\initparam \gets$ Random Initialization
	\For{$d\gets 1,K$} \Comment{$K$ is the number of meta-training steps}
		\State $\trainset_c, \testset_c \leftarrow$ Random scenario from $\metatrainset$
		\State $\recomparam^{(0)}_c\gets\initparam,T\gets 0$
		\For{$t\gets 1,T_{max}$}
			\State $B^{(t)} \gets$ Random batch from $\trainset_c$
			\State $\trainloss^{(t)}\gets\mathcal{L}(B^{(t)};\recomparam^{(t-1)}_c)$
			\State $p^{(t)}\gets M_s(\trainloss^{(t)}, ||\nabla_{\theta^{(t-1)}_c}{\trainloss^{(t)}}||_{2};\stopparam)$ \Comment{\Cref{eqn:early-stop}}
			\State $s^{(t)}\sim\mathrm{Bernoulli}(p^{(t)})$ \Comment{Randomly decide to stop or not}
			\If{$s^{(t)}=1$}
			\State Break;
			\EndIf
			\State $\recomparam^{(t)}_c\gets$ Update $\recomparam^{(t-1)}_c$ according to \Cref{eqn:lstm_update,eqn:update_control}
			\State $T\gets T+1$
		\EndFor
		\State $\testloss\gets\mathcal{L}(\testset_c;\recomparam^{(T)}_c)$
		\State Update $\updateparam$, $\initparam$ using $\nabla_{\nupdateparam}\testloss$, $\nabla_{\ninitparam}\testloss$
		\State $d\stopparam\gets 0$
		\For{$j\gets 1,T$}
		\State $d\stopparam\gets d\stopparam+\big(\testloss-\mathcal{L}(\testset_c;\recomparam^{(j)}_c)\big)\nabla_{\stopparam}\ln(1-p^{(j)})$
		\EndFor
		\State Update $\stopparam$ using $d\stopparam$ \Comment{\Cref{eqn:reinforce}}
	\EndFor
\end{algorithmic}
\end{algorithm}

In the scenario-specific learning, the recommender parameters $\recomparam_c$ are learned from the training set $\trainset_c$. We summarize the following three challenges in the learning:

\begin{itemize}[leftmargin=*]
	\item  How should the parameters $\recomparam_c$ be initialized? Randomly initialized parameters can take a long time to converge and often lead to overfitting in the few-shot setting. 
	\item How should the parameters $\recomparam_c$ be updated w.r.t the loss function? Traditional optimization algorithms rely on carefully tuned hyperparameters to converge to a good solution. Optimal hyperparameters on different scenarios may vary a lot.
	\item When should the learning process stop? In few-shot setting, learning too much from a small training set can lead to overfitting and hurt the generalization performance. 
\end{itemize}
These challenges are often solved by experts manually in traditional machine learning settings. Instead, we propose \smodel which can automatically learn to control the learning process from end to end, including parameter initialization, update strategy and early-stop policy. In the following part, we will introduce how the meta learner controls the three parts of scenario-specific learning and how the meta learner is trained on the meta-training set $\metatrainset$.

\subsubsection{Parameter Initialization}
At the beginning of scenario-specific learning, the recommender parameters are initialized as $\recomparam^{(0)}_c$.
Traditionally, the parameters of a neural network are initialized by randomly sampling from a normal distribution or uniform distribution. Given enough training data, the randomly initialized parameters can usually converge to a good local optimum but may take a long time. In the few-shot setting, however, random initialization combined with limited training data can lead to serious overfitting, which hurts the ability of the trained recommender to generalize well. Instead, following \cite{MAML:Finn}, we initialize the recommender parameters from the global initial values shared across different scenarios. These initial values are considered as one of meta learner's parameters, denoted as $\initparam$. Suitable initial parameters may not perform well on a specific scenario, but can adapt quickly to new scenarios given a small amount of training data.
 
\subsubsection{Update Strategy}
At each step $t$, a batch $B^{(t)}=\{u_k, i_k, i_k^-\}_{k=1}^N$ with the fixed size $N$ is sampled from $\trainset_c$, where $(u_k, i_k)\in\trainset_c$ and $i_k^-$ is sampled from $\mathcal{I}$ such that $(u_k, i_k^-)\notin\trainset_c$. Then the previous parameters $\recomparam^{(t-1)}_c$ are updated to $\recomparam^{(t)}_c$ according to $\mathcal{L}^{(t)}$, the loss on $B^{(t)}$:

\begin{align}
	\mathcal{L}^{(t)} &= \sum_{(u_k,i_k,i_k^-\in B^{(t)})}\frac{\ell(u_k,i_k,i_k^-;\recomparam^{(t-1)}_c)}{|B^{(t)}|},
\end{align}

\noindent where $\ell$ is the loss function in \Cref{eqn:loss_function}.

The most common method to update parameters of a neural network is stochastic gradient descent (SGD) \cite{SGD:robbins1951}. In SGD, the parameters $\recomparam_c$ are updated as:
\begin{equation}
	\recomparam^{(t)}_c=\recomparam^{(t-1)}_c-\alpha\nabla_{\nrecomparam^{(t-1)}_c}\trainloss^{(t)},
\end{equation}
where $\alpha$ is the learning rate. There are many variations based on SGD, such as Adam \cite{adam:kingma} and RMSprop \cite{RMSprop:tielemanH12}, both of which adjust the learning rate dynamically.

Although hand-crafted optimization algorithms have gained success in training deep networks, they rely on a large amount of training data and carefully selected hyperparameters to converge to a good solution. In few-shot learning, the inappropriate learning rate can easily lead to being stuck in a poor local optimum. Moreover, optimal learning rates in different scenarios can differ significantly. Instead, we extend the idea of learning the optimization algorithm in \cite{Meta-LSTM:Ravi}  to learn an \emph{update controller} for parameters in $\recomparam_c$, implementing a more flexible update strategy than hand-crafted algorithms. The update strategy for $\recomparam_c$ is:

\begin{equation}
	\label{eqn:lstm_update}
	\recomparam^{(t)}_c=\myvec{\forgetgate}^{(t)}\odot\recomparam^{(t-1)}_c - \myvec{\inputgate}^{(t)}\odot\nabla_{\nrecomparam^{(t-1)}_c}\trainloss^{(t)},
\end{equation}

\noindent where $\myvec{\inputgate}^{(t)}$ is the \emph{input gate}, similar to the learning rate in SGD; and $\myvec{\forgetgate}^{(t)}$ is the \emph{forget gate}, which can help the meta learner quickly "forget" the previous parameters to leave a poor local optimum.

Since the optimization process is a sequential process, in which the historical information matters, we use LSTM \cite{LSTM:HochreiterS97} to encode historical information and issue input gates and forget gates: 
\ignore{
In the learning process, each scalar in $\recomparam$ independently updates its cell state and hidden state according to its value, gradient, and the loss, and its input gate and forget gate are computed from its hidden state. 
}

\begin{equation}
\label{eqn:update_control}
\begin{split}
	\myvec{h}^{(t)}_u, \myvec{c}^{(t)}_u &= \mathrm{LSTM}([\nabla_{\nrecomparam^{(t-1)}_c}{\trainloss^{(t)}},\trainloss^{(t)},\recomparam^{(t-1)}_c], \myvec{h}_u^{(t-1)}, \myvec{c}_u^{(t-1)}),\\
	\myvec{\forgetgate}^{(t)}&=\sigma(\mymat{W}_F\myvec{h}_u^{(t)}+\myvec{b}_F),\\
	\myvec{\inputgate}^{(t)}&=\sigma(\mymat{W}_I\myvec{h}_u^{(t)}+\myvec{b}_I),
\end{split}
\end{equation}

\noindent where $\mathrm{LSTM}(\cdot,\cdot,\cdot)$ represents one-step forward in standard LSTM, $\myvec{h}_u^{(t)}$ and $\myvec{c}_u^{(t)}$ are hidden state and cell state of the update controller at step $t$, and $\sigma(\cdot)$ is the sigmoid function. The parameters of the updated controller, denoted as $\updateparam$, include $\{\mymat{W}_F,\myvec{b}_F,\mymat{W}_I,\myvec{b}_I\}$ as well as LSTM related parameters. Different parameters in $\recomparam_c$ correspond to different update controllers. In this way, different learning strategies can be applied.

\subsubsection{Early-Stop Policy}
In machine learning, overfitting is one of the critical problems that restrict the performance of models. When the model's representation ability exceeds the complexity of the problem, which is often the case for neural networks, the model might fit the sampling variance and random noise in the training data and get poor performance on the testing set. Regularization tricks like $L_2$ regularizer or dropout\cite{Dropout:Srivastava2014} are often applied to limit the model's complexity. Another common trick is early-stop: the learning process is stopped when the training loss stops descending or the performance on the validation set begins to drop. 

In few-shot setting, as we found, regularizers cannot prevent overfitting to the small training set. Also, as the size of the training set is too small, the validation set divided from the training set cannot provide a precise estimation of generalization ability. To overcome the drawback of hand-crafted stop rules, we propose to learn the stop policy with a neural network, which we call \emph{stop controller} $M_s$. To balance exploitation and exploration, we apply a stochastic stop policy, in which at step $t$, the learning process stops with probability $p^{(t)}\ignore{=M_s(\trainloss^{(t)}, ||\nabla_{\theta^{(t-1)}_c}{\trainloss^{(t)}}||_{2})}$, which is predicted by $M_s$. Similar to the update controller, $M_s$ is an LSTM that can encode historical information:

\begin{equation}
\label{eqn:early-stop}
\begin{split}
	\myvec{h}_s^{(t)}, \myvec{c}_s^{(t)} &= \mathrm{LSTM}([\trainloss^{(t)}, ||\nabla_{\nrecomparam^{(t-1)}_c}{\trainloss^{(t)}}||_{2}], \myvec{h}_s^{(t-1)}, \myvec{c}_s^{(t-1)}),\\
	p^{(t)} &= \sigma(\myvec{W}_s\myvec{h}_s^{(t)}+b_s),
\end{split}
\end{equation}

\noindent where $||\nabla_{\nrecomparam^{(t-1)}_c}{\trainloss^{(t)}}||_2$ is $L_2$-norm of gradients at step $t$, and $h^{(t)}_s$ and $c^{(t)}_s$ are the hidden state and the cell state of the stop controller at step $t$. The parameters of the stop controller, denoted as $\stopparam$, include $\{\mymat{W}_s,\myvec{b}_s\}$ as well as LSTM related parameters.

\subsubsection{Training of Meta Learner}
\label{subsubsec:train_meta_leaner}
The objective of the meta learner is to minimize the expected loss on the testing set after scenario-specific learning on the training set, as is described in \Cref{eqn:objective,eqn:loss_function,eqn:test_loss}. The parameters $\metaparam$ of the meta learner include shared initial parameters $\initparam$, parameters of the update controllers $\updateparam$ and parameters of the stop controller $\stopparam$.

The gradients of $\initparam$ and $\updateparam$ with respect to the meta-testing loss $\mathcal{L}_\omega(\testset_c|\trainset_c)$, which we call \emph{meta-gradient}, can be computed with back-propagation.
%\begin{equation}
%\begin{split}
%	&\recomparam_c = \myvec{\forgetgate}_0\odot\recomparam_0-\myvec{\inputgate}_0\odot\nabla_{\recomparam_0}\trainloss_0, \\
%	&\nabla_{\initparam}\testloss \\
%%	=&\nabla_{\theta_c}\testloss\cdot\nabla_{\initparam}\theta_c,\\
%	=&\nabla_{\recomparam_c}\testloss\cdot[\nabla_{\recomparam_0}\myvec{\forgetgate}_0\odot\recomparam_0+\myvec{\forgetgate}_0-\myvec{\inputgate}_0\odot H_{\recomparam_0}(\trainloss_0)-\nabla_{\recomparam_0}\myvec{\inputgate}_0\odot\nabla_{\recomparam_0}\trainloss_0] \\
%	&\nabla_{\updateparam}\testloss
%	=\nabla_{\recomparam_c}\testloss\cdot[\nabla_{\updateparam}\myvec{\forgetgate}_0\odot\recomparam_0-\nabla_{\updateparam}\myvec{\inputgate}_0\odot\nabla_{\recomparam_0}{\trainloss_0}].
%\end{split}
%\end{equation}
However, the meta-gradient may involve higher-order derivatives, which are quite expensive to compute when $T$ is large. Therefore in MAML\cite{MAML:Finn}, they only take one-step gradient descent. Instead, our model can benefit from flexible multi-step gradient descent. We ignore higher-order derivatives in the meta-gradient by taking the gradient of the recommender parameters, $\nabla_{\nrecomparam^{(t-1)}_c}\trainloss^{(t)}$, as independent of $\initparam$ and $\updateparam$. With the gradients, we can optimize $\initparam$ and $\updateparam$ with normal SGD.

Since the relation between $\stopparam$ and $\recomparam_c$ is discrete and stochastic, it is impossible to take direct gradient descent on $\stopparam$. We use stochastic policy gradient to optimize $\stopparam$. Given one learning trajectory based on stop controller parameters $\stopparam$: $\recomparam^{(0)}_c,\recomparam^{(1)}_c,\recomparam^{(2)}_c,\cdots,\recomparam^{(T)}_c$, We define the immediate reward at step $t$ as the loss decrease of one-step update: $r^{(t)} = \mathcal{L}(\testset_c;\recomparam^{(t-1)}_c) - \mathcal{L}(\testset_c;\recomparam^{(t)}_c)$. Then the accumulative reward at $t$ is:

\begin{equation}
	Q^{(t)}=\sum_{i=t}^T{r^{(t)}}=\mathcal{L}(\testset;\recomparam^{(t-1)}_c)-\mathcal{L}(\testset;\recomparam^{(T)}_c),
\end{equation}

\noindent which is the loss decrease from step $t$ to the end of learning \ignore{subtracted by the accumulative penalty}. According to the REINFORCE algorithm\cite{REINFORCE:Williams92}, we can update $\stopparam$ as:

\begin{equation}
	\label{eqn:reinforce}
	\stopparam\gets\stopparam+\gamma\sum_{t=1}^T{Q^{(t)}\nabla_{\nstopparam}\mathrm{ln}M_s(\trainloss^{(t)}, ||\nabla_{\nrecomparam^{(t-1)}_c}{\trainloss^{(t)}}||_{2};\stopparam)},
\end{equation}

\noindent where $\gamma$ is the learning rate for $\stopparam$. The detailed training algorithm for the \smodel learner is described in \Cref{algo:meta_train}.
\section{Experiment}
\label{sec:experiment}
In this section, we detail the experiments to evaluate the proposed \smodel in the few-shot scenario-aware recommendation, including details of the datasets, competitors, experimental settings and comparison results. We will also delve deeper to analyze how \smodel helps the recommender network achieves better results with process sensitivity analyses. Finally, we analyze how the architecture of the recommender influences performance.

\subsection{Experimental Settings}
\subsubsection{Datasets}\label{subsubsec:dataset} 
Current available open datasets for context-aware recommender systems are mostly of small scale with very sparse contextual information \cite{ContextDataset:IlarriLH18}. We evaluate our method on two public datasets and one newly released large scenario-based dataset from Taobao\footnote{Downloadable from \url{https://tianchi.aliyun.com/dataset/dataDetail?dataId=9716}.}. 
The first dataset is the Amazon Review dataset\cite{AmazonReview:HeM}, which contains product reviews and metadata from Amazon. We keep the reviews with ratings no less than three as the positive user-item interactions and take interactions in different categories as different scenarios. 
Our second dataset is the Movielens-20M \cite{Movielens:HarperK16} dataset, with rating scores from the movie recommendation service Movielens.  Following \cite{NeuralMF:HeLZNHC17}, we transform the explicit ratings into implicit data, where all the movies the user has rated are taken as positive items for the user. The tags of movies are used as scenarios. 
In each dataset, we select scenarios with less than 1,000 but more than 100 items/movies as few-shot tasks with enough interactions for evaluation. 
Our third dataset is from the click log of Cloud Theme, which is a crutial recommendation procedure in the Taobao app. Different themes correspond to different scenarios of purchase, e.g., \emph{"what to take when traveling"} \emph{"how to dress up yourself on a party"} \emph{"things to prepare when a baby is coming"}. In each scenario, a collection of items in related categories is displayed, according to the scenario as well as the user's interests. The dataset includes more than 1.4 million clicks from 355 different scenarios in a 6-days promotion season, with one-month purchase history of users before the promotion started. Table \ref{tab:dataset_statistics} summarizes the statistics of three datasets.  
\subsubsection{Baselines} We select state-of-the-art baselines in item recommendation in the same domain or cross-domains which can be broadly divided into the following categories:
\begin{description}
\item [Heuristic:] ItemPop: Items are ranked according to their popularity in specific scenarios, judged by the number of interactions in corresponding training sets. This is a non-personalized baseline in item recommendation\cite{BPR:RendleFGS09}.

\item [General Domain:] NeuMF: The Neural Collaborative Filtering\cite{NeuralMF:HeLZNHC17} is a state-of-the-art item recommendation method. It combines the traditional MF and MLP in neural networks to predict user-item interactions. This method is not proposed for cross-domain recommendation, and we train a single model on all the scenarios.
\item [Shallow Cross-Domain:] 
\begin{enumerate*}
	\item CDCF: The Cross-Domain Collaborative Filtering\cite{CDCF:LoniSLH14} is a simple method based on Factorization Machines \cite{FM:Rendle10} for cross-domain recommendation. By factorizing the interaction data of shared users in the source domain, it allows extra information to improve recommendation in a target domain;
	\item CMF: Collective Matrix Factorization\cite{CMF:SinghG08} is a matrix factorization method for cross-domain recommendation. It jointly learns low-rank representations for a collection of matrices by sharing common entity factors, which enables the knowledge transfer across domains. We use it as a shallow cross-domain baseline, with users partially overlapping in different domains.
\end{enumerate*}
\item [Deep Cross-Domain:]
\begin{enumerate*}
	\item EMCDR: The Embedding and Mapping framework for Cross-Domain Recommendation \cite{EMCDR:Man} consists of two steps. Step 1 is to learn the user and item embeddings in each domain with latent factor models. Step 2 is to learn the embedding mapping between two domains with a multilayer perceptron from the shared users/items in two domains; 
	\item CoNet: The Collaborative Cross Networks\cite{CoNet:HuZY18} enables dual knowledge transfer across domains by introducing cross connections from one base network to another and vice versa. 
%\ignore{Let $\myvec{z}_s$ and $\myvec{z}_t$ be the representations of the $l$-th hidden layers in the source and target networks respectively and $\tilde{\myvec{z}}_s$ and $\tilde{\myvec{z}}_t$ be the inputs to the activation functions of $l$+1-th hidden layers:
%		
%	\begin{equation}
%		\begin{split}
%			\tilde{\myvec{z}}_s &= \mymat{W}_s\myvec{z}_s+\mymat{H}\myvec{z}_t,\\
%			\tilde{\myvec{z}}_t &= \mymat{W}_t\myvec{z}_t+\mymat{H}\myvec{z}_s,
%		\end{split}
%	\end{equation}
%	
%	\noindent where $\mymat{W}_s$ and $\mymat{W}_t$ are weight matrices and $\mymat{H}$ is the transformation matrix that controls the knowledge transfer. }
\end{enumerate*}
\end{description}
\begin{table}[!ht]
	\centering
	\caption{Statistics of the Datasets. \#Inter. denotes the number of user-item interactions and \#Scen. denotes the number of scenarios we use as few-shot tasks.}
	\begin{tabular}{crrrr}
		\toprule
		Dataset & \#Users & \#Items & \#Inter. & \#Scen.\\
		\midrule
		Amazon & 766,337 & 492,505 & 17,523,124 & 1,289\\
		Movielens & 138,493 & 27,278 & 20,000,263 & 306\\
		Taobao & 775,603 & 1,452,525 & 5,717,835 & 355 \\
		\bottomrule
	\end{tabular}
	\label{tab:dataset_statistics}
\end{table}

 \begin{table*}[!ht]
	\caption{The top-N recall results on test scenarios.}
	\label{tab:performance}
	\centering
		{
	\begin{tabular}{l| *{3}{r} | *{3}{r} | *{3}{r}}
	\toprule
	\multicolumn{1}{c|}{} & 
	\multicolumn{3}{c|}{\textbf{Amazon}} &
	\multicolumn{3}{c|}{\textbf{Movielens}} &
	\multicolumn{3}{c}{\textbf{Taobao}}
	\\\cline{2-10}
	\multicolumn{1}{l|}{Method} & 
	\multicolumn{1}{r}{Recall@10} &
	\multicolumn{1}{r}{Recall@20} &
	\multicolumn{1}{r|}{Recall@50} &
	\multicolumn{1}{r}{Recall@10} &
	\multicolumn{1}{r}{Recall@20} &
	\multicolumn{1}{r|}{Recall@50} &
	\multicolumn{1}{r}{Recall@20} &
	\multicolumn{1}{r}{Recall@50} &
	\multicolumn{1}{r}{Recall@100} 
	\\\midrule
	NeuMF & 24.55 & 35.65 & 55.19 & 31.67 & 51.30 & 84.98 & 25.64 & 42.31 & 58.84\\
%	Embedding & - & - & - & 12.86 & 20.99 & 37.46\\
	\midrule
	ItemPop & 26.86 & 32.65 & 50.42 & 39.65 & 54.32 & 78.12 & 18.25 & 28.57 & 32.44\\
	CDCF & 10.27 & 16.72 & 32.79 & 29.19 & 41.04 & 65.75 & 9.81 & 20.93 & 32.14\\
	CMF & 27.64 & 38.31 & 55.24 & 29.80 & 46.91 & 74.37 & 9.16 & 16.47 & 29.31\\
	EMCDR & 31.71 & 42.14 & 58.73 & 43.55 & 60.89 & 83.54 & 20.43 & 31.52 & 45.67\\
	CoNet & 30.17 & 41.57 & 56.06 & 46.62 & 63.61 & 87.06 & 20.27 & 31.48 & 44.53\\
	\midrule
%	No finetune & 40.43 & 58.12 & 87.34 & 19.04 & 27.97 & 44.55\\
%	Meta-MAML  & 33.45 & 45.60 & 63.63 & 21.36 & 30.99 & 47.97\\
	\model & \textbf{34.39} & \textbf{46.53} & \textbf{64.28} & \textbf{47.79} & \textbf{66.02} & \textbf{89.07} & \textbf{27.11} & \textbf{44.10} & \textbf{59.98}\\
	\midrule
	Improve & 8.35 & 10.42 & 9.45 & 2.51 & 3.79 & 2.31 & 5.73 & 4.23 & 1.90\\
	\bottomrule
	\end{tabular}
	}
\end{table*}

More details in experimental settings can be found in \Cref{sec:appendix}.

\subsection{Performance Comparison}
In this subsection, we report comparison results and summarize insights.  \Cref{tab:performance} shows the results on the three datasets concerning top-N recalls \cite{Caser:TangW18}. We can see that \smodel achieves the best results throughout three datasets, compared to both shallow cross-domain models (CMF and CDCF) and deep cross-domain models (EMCDR and CoNet). Involving the extracted scenario information, \smodel also performs better compared to the general recommendation (NeuMF) and the heuristic method ItemPop. 

For the  Amazon dataset, \smodel gives $9.41\%$ improvements on average compared with the best baseline (EMCDR), and achieves $16.47\%$ improvements in terms of Recall@50 compared to the non-scenario-aware NeuMF, which demonstrates the benefits of combining scenario information. Among the cross-domain baselines, neural cross-domain methods are slightly better than the shallow cross-domain methods.

For the Movielens dataset, \smodel achieves $2.87\%$ improvements on average compared to the best baseline (CoNet). It is a phenomenon common in Amazon and Movielens that domain-specific methods can perform better than the left competitors, showing that the user interests relatively concentrate in a specific scenario. Also, we can find that the deep cross-domain methods can outperform shallow cross-domain methods by improvements over $10\%$. The performance of NeuMF, which is not designed for cross-domain recommendation, is also superior to CMF and CDCF, showing the power of deep learning in recommendation.
 
For the Taobao dataset, \smodel achieves $3.95\%$ performance lifts on average. For this dataset, NeuMF performs best among all the baselines. The reason might be that this dataset is from the click log in the real-world recommendation and clicks often contain more noise than purchase or ratings. The performances of baselines that adapt to the scenario in a naive way might be detrimental by overfitting in the few-shot setting. Above all, \smodel can still outperform NeuMF with a flexible learning strategy learned by the proposed meta learner.
 
Note that the relative improvements by taking account of scenario information vary among three datasets. For Amazon, most scenario-specific baselines can perform better than the general domain baseline, while for the Taobao dataset, most scenario-specific baselines cannot outperform NeuMF. It implies that the usefulness of scenario information varies, and \smodel can dynamically adapt to different datasets' requirements.

\subsection{Process Sensitivity Analysis}
In this subsection, we analyze how the \smodel works to control the scenario-specific learning process. Due to the space limit, we only illustrate results from Amazon and performance patterns are the same from the other two datasets. 
 
\subsubsection{Impact of Different Parts}
\begin{figure*}[!ht]
\includegraphics[width=\textwidth]{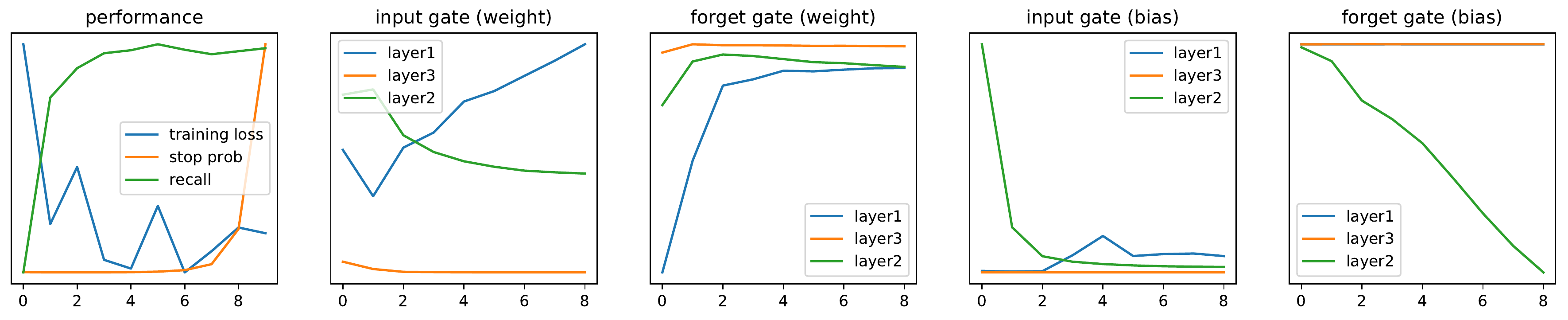}
\caption{The visualization of the learning process. Left most is the training loss, stop probability and recall on the testing set during the learning process. The other four figures are average input gates and forget gates of weights and biases in the hidden layers.}
\label{fig:learning_process}
\end{figure*}

\begin{figure*}[!htb]
\includegraphics[width=0.9\textwidth]{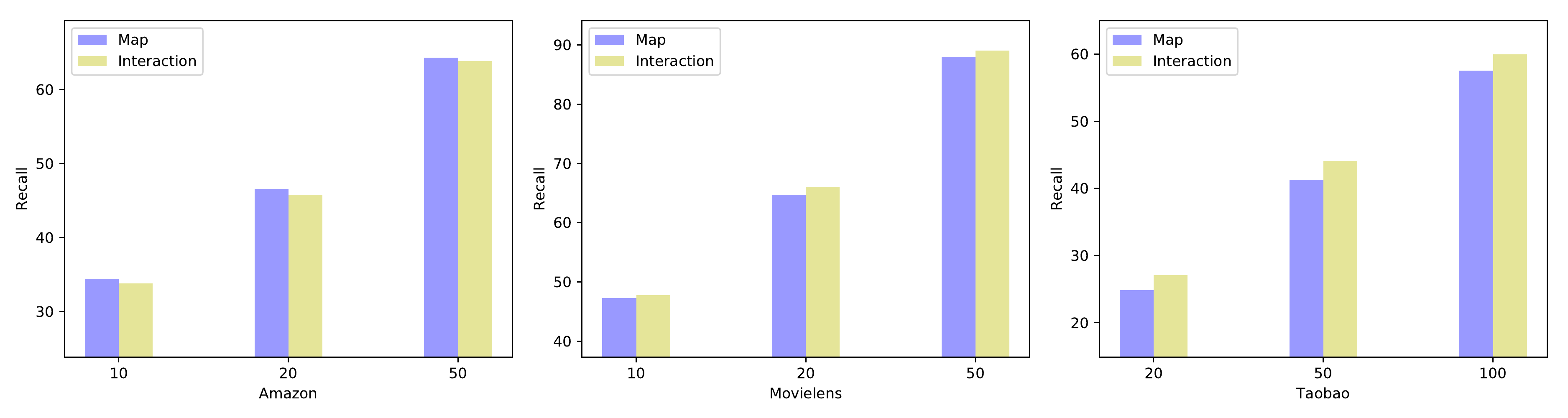}
\caption{Impact of the recommender architecture on the Amazon(left), Movielens(middle) and Taobao(right) datasets. }
\label{fig:recommender_architecture}
\end{figure*}%

First, we analyze the contributions of three parts of meta learner described in \Cref{subsec:meta_learner}. Specifically, we compare the performances of the following variations:
\begin{itemize}
	\item \emph{Complete:} The complete meta learner that controls parameter initialization, update strategy, and early-stop policy.
	\item \emph{RandInit:} Parameters of the recommender are randomly initialized. The update strategy and early-stop policy remain unchanged.
	\item \emph{FixedLr:} Parameters of the recommender are updated by standard SGD, with fixed learning rate 0.01. Parameter initialization and early-stop policy remain unchanged.
	\item \emph{FixedStep:} The step of scenario-specific learning is fixed at $20$, while parameter initialization and update strategy remain consistent with the complete method.
\end{itemize}

The performance comparison is listed in \Cref{tab:meta_learner_part}. We can see that the complete model can significantly outperform the three weakened variations, indicating that three parts of the meta learner all help to improve the final performance. Among the three variations, the random initialization hurts the performance most. When the parameters are randomly initialized, the recommender has to learn from a random point each time and the learning will take more steps and more easily be stuck in a local optimum. The effect of removing early-stop policy is relatively small. The reason might be that when the learning process is too long, the update controller can still give low input gates to avoid overfitting.
\begin{table}[!ht]
	\caption{Impact of different parts in meta learner on Amazon dataset}
	\label{tab:meta_learner_part}
	\centering
	\begin{tabular}{c c c c}
		\toprule
		Method & Recall@10 & Recall@20 & Recall@50\\
		\midrule
		RandInit & 33.12 & 45.44 & 63.35\\
		FixedLr & 33.56 & 45.90 & 63.86\\
		FixedStep & 33.84 & 46.13 & 64.02\\
		Complete & \textbf{34.39} & \textbf{46.53} & \textbf{64.28}\\
		\bottomrule
	\end{tabular}
\end{table}

\subsubsection{Case Study: the learning process learned by the meta learner}

To further analyze the learning process the meta learner learned automatically, we select a typical scenario on Amazon and visualize its learning process. \Cref{fig:learning_process} shows the training loss, stop probability(predicted by the stop controller), recall on the test set(not visible to the meta learner) and the input and forget gates(predicted by the update controller). From the leftmost figure, we can see that as the training loss ceases to drop, the stop probability rises dramatically, finally leading to the end of the learning process. In fact we can find that the recall on the test set has been stable and further training might lead to overfitting. From the visualization of input gates and forget gates, different update strategies are applied for different layers and different parameters. For different layers, we can find that for the last hidden layer, the input gates remain low and the forget gates remain high, which indicates its parameters remain stable during training. Among different parameters, the weight matrices mainly change at the beginning, and the bias vectors mainly change at the end. This strategy might help to quickly adjust the parameters by updating a part of the parameters at a time. Also, updating the bias vectors can be a minor complement to the updated weight matrices.

\subsection{Analysis of Recommender Architecture}
\label{subsec:recommender_architecture}

In this subsection, we analyze how the architecture of the recommender influences the performance. We compare the following two architectures for the recommender:
\begin{itemize}
	\item \emph{Mapping Module:} The user and item embeddings are mapped by two multilayer perceptrons respectively, and the final score is computed as the dot product of the mapped user and item embeddings. This is the architecuture of EMCDR.
	\item \emph{Interaction Module:} User and item embeddings are concatenated and a multilayer perceptron computes the final score from the concatenated vector. This is the architecture used by NeuMF and CoNet.
\end{itemize}

We compare the performance of two architectures with the same meta learner. The results on three datasets are shown in \Cref{fig:recommender_architecture}. In general, we can see that in general the interaction module can perform better than the mapping module, because the interaction module can represent the complicated interactions between users and items, while the mapping module only maps the user and item embeddings independently. On Taobao, the relative improvement of interaction modules can reach $6.71\%$. However, on Amazon, the mapping module slightly outperforms the interaction module. This implies that the optimal recommender architecture may differ among datasets. It's also noted that in practice the mapping module can be more efficient because given the recommender, the mapped embeddings can be pre-computed and the matching can be optimized with advanced data structures.  

%\begin{table}
%	\begin{tabular}{c|c|c|c}
%		\toprule
%		Method & Recall@10 & Recall@20 & Recall@50\\
%		\midrule
%		item & 21.96 & 31.92 & 49.12\\
%		user & 21.62 & 31.49 & 48.97\\
%		user+item & 21.54 & 31.11 & 48.23\\
%		user+item+bias & 20.84 & 30.68 & 48.05 \\
%		\bottomrule
%	\end{tabular}
%\end{table}

\section{Conclusion}
\label{sec:conclusion}

In this work, we explored few-shot learning for recommendation in the scenario-specific setting. We proposed a novel sequential scenario-specific framework for recommender systems using meta learning, to solve the cold-start problem in some recommendation scenarios. \smodel can automatically control the learning process to converge to a good solution and avoid overfitting, which has been the critical issue of few-shot learning. Experiments on real-world datasets demonstrate the effectiveness of the proposed method, by comparing with shallow/deep, general/scenario-specific baselines. In the future, we will automate the architecture design of recommenders. In our experiments, we found that the performance of the same architecture might differ in different datasets. By learning to choose the optimal recommender architecture, the performance of \smodel can be further improved. 

\begin{acks}
Jie Tang and Hongxia Yang are the corresponding authors of this paper. 
The work is supported by the
NSFC for Distinguished Young Scholar 
%jie tang
(61825602),
Tsinghua University Initiative Scientific Research Program,
and a research fund supported by Alibaba.
\end{acks}

%
% The next two lines define the bibliography style to be used, and the bibliography file.
\balance
\bibliographystyle{ACM-Reference-Format}
\bibliography{reference}

\clearpage
\nobalance
\appendix

\section{Appendix}
\label{sec:appendix}

In this section, we first give the deployment description and implementation notes of our proposed models. The implementation notes and parameter configurations of compared methods are then given. Finally, we introduce the experiment settings in three datasets.
\vspace{-0.1in}
\subsection{Deployment}
Deployment is at the Guess You Like session, the front page of the Mobile Taobao; and the illustration video can also be watched from the link\footnote{\url{https://youtu.be/TNHLZqWnQwc}}.
\vspace{-0.1in}
\subsection{Implementation Notes}
The architecture of the recommender network is designed as three hidden layers and one output layer. The embedding size is set to 128 on Amazon and Alibaba Theme and 64 on Movielens. The number of hidden units in each layer is half of that in the last layer. The hidden size of the update LSTM is set to 16. The hidden size of the stop LSTM is set to 18. For the input of the update controllers and stop controller, we use the preprocessing trick in \cite{GradientByGradient:AndrychowiczDCH16}:
\begin{equation}
	x\rightarrow
	\begin{cases}
		(\frac{\log(|x|)}{p}, \mathrm{sgn}(x)) & \text{if $|x|\ge e^{-p}$}\\
		(-1, e^px) & \text{otherwise}
	\end{cases}
\end{equation}
\noindent where $p=10$ in our experiment. \smodel is trained by standard SGD with learning rate $0.0001$ and weight decay 1e-5,  implemented with PyTorch\footnote{\url{https://pytorch.org/}} 1.0 in Python 3.6 and runs on a single Linux server with 8 NVIDIA GeForce GTX 1080.
\vspace{-0.1in}
\subsection{Compared Methods}
\label{subsec:parameters}
 
\subsubsection{Code} The code of NeuMF is provided by the author\footnote{\url{http://github.com/hexiangnan/neural_collaborative_filtering}}. We simply adapt the input and evaluation modules to our experimental settings. For CMF, we adpot the implementation of LIBMF\footnote{\url{https://www.csie.ntu.edu.tw/~cjlin/libmf/}}. For CDCF, we dapot the official libFM implementation\footnote{\url{http://www.libfm.org}}. The codes of EMCDR and CoNet are not published and we implement them with PyTorch.
\vspace{-0.1in}
\subsubsection{Parameter Configuration} We do grid search over the hyperparameters of CDCF, CMF, EMCDR, and CoNet, with the meta-training set. The search space for CDCF and CMF is learning rate $(0.1, 0.03, 0.001)$, learning step $(10, 100, 200)$ and factor dimension $(8, 16, 32)$. The search space for EMCDR is learning rate\\$(0.1, 0.01, 0.001, 0.0001)$ and learning step $(10, 100, 1000)$. The search space for CoNet is learning rate $(0.01, 0.001)$, learning step $(10, 50, 100, 500)$ and sparse ratio $\lambda$ $(0.1,0.01,0.001)$. For NeuMF, we use the pre-training suggested by the author, with embedding size of MF model set to 8 and MLP layers set to $[64,32,16,8]$. 
\vspace{-0.1in}
\subsection{Experiment Setting}
As we mention in \Cref{subsubsec:dataset}, it is challenging to find large-scale public datasets to evaluate context-aware recommender systems. Therefore we evaluate our method on two public datasets in the cross-domain setting. We also use one scenario-based dataset in Taobao.

On the Amazon Review\footnote{\url{http://jmcauley.ucsd.edu/data/amazon}} dataset, we take user purchases in different leaf categories as different scenarios. To maximize the difference between the source domain and the target domain and fully evaluate the model's ability to learn user behaviors in new scenarios, we select the leaf categories in different first-order categories as the source domain, meta-training set and meta-testing set. The specific split of first-order categories is displayed in \Cref{tab:split_amazon}. 

On the Movielens-20M\footnote{\url{https://grouplens.org/datasets/movielens/}} dataset, we take the movie tags in the Tag Genome included in the dataset as scenarios. The tag genome is a data structure that contains tag relevance scores for movies that are computed based on user-contributed content. Tags with relevance scores higher than 0.8 to a movie are considered as tags of the movie. We use the movies without tags as the source domain, and randomly divide the tags into the meta-training set and meta-testing set.

On both datasets, we select the scenarios with 100-1000 items/movies as the few-shot tasks with enough interactions for evaluation. Unlike traditional cross-domain settings in which the interactions on different domains are simply concatenated, we are more interested in the cold-start scenarios, on which most users have no previous interactions. Therefore on each scenario, 64 interactions of shared users are used as $\trainset_c$ and all others are used for evaluation. We use all the interactions on the source domain to generate the user embeddings and the interactions of the users only appearing on the scenario to generate the item embeddings, with Matrix Factorization\cite{MFforRS:KorenBV09}. For cross-domain baselines, each scenario is considered as a target domain. The training data is all the interactions on the source domain, the few-shot interactions of shared users $\trainset_c$, and all the interactions of the users only appearing on the target domain. Therefore, all the comparative methods use the same training data and evaluation set, and the fairness of comparison is guaranteed. 

On the Taobao Themes\footnote{\url{https://tianchi.aliyun.com/dataset/dataDetail?dataId=9716}.} dataset, we take the different themes as different scenarios and all the clicks are considered as positive user-item interactions. We use the one-month purchase history of users before the click log as the source domain and randomly divide the scenarios into the meta-training set and meta-testing set. As this dataset is quite sparse, we use the general user and item embeddings generated from attributes and interactions in Taobao by GraphSage\cite{GraphSage:HamiltonYL17}. This embeddings are also used by NeuMF, EMCDR and CoNet. 

%\begin{table}[!ht]
%	\caption{Dataset split on Amazon}
%	\begin{tabular}{c|c|c}
%		\toprule
%		source domain & meta-training & meta-test\\
%		\midrule
%		Books & Tools and Home Improvement\\
%		Electronics &  CDs and Vinyl\\
%		Movies and TV & Kindle Store\\
%		Clothing, Shoes and Jewelry & Beauty\\
%		Home and Kitchen & Pet Supplies\\
%		Sports and Outdoors\\
%		Health and Personal Care\\
%		Toys and Games\\
%		Apps for Android\\
%		Grocery and Gourmet Food\\
%		\bottomrule
%	\end{tabular}
%\end{table}

\begin{table}[!ht]
	\caption{Dataset split on Amazon}
	\label{tab:split_amazon}
\vspace{-0.1in}
	\begin{tabular}{c|c}
	\toprule
		Domain & First-order Category\\ 	
		\midrule
		Source Domain & \tabincell{c}{Books, Electronics, Movies and TV\\Clothing, Shoes and Jewelry, Home and Kitchen\\ Sports and Outdoors, Health and Personal Care\\Toys and Games, Apps for Android\\Grocery and Gourmet Food}\\
		\hline
		Meta-training & \tabincell{c}{Beauty, CDs and Vinyl, Kindle Store\\Tools and Home Improvement, Pet Supplies} \\
		\hline
		Meta-testing & \tabincell{c}{Digital Music, Musical Instruments\\ Video Games, Cell Phones and Accessories\\Baby, Automotive, Office Products\\Amazon Instant Video, Patio, Lawn and Garden}\\
		\bottomrule
	\end{tabular}
\end{table}

\end{document}